\documentclass[letterpaper,journal]{IEEEtran}

\IEEEoverridecommandlockouts
\usepackage{cite}
\usepackage{amsmath,amssymb,amsfonts}
\usepackage{algorithmic}
\usepackage{graphicx}
\usepackage{textcomp}
\usepackage{xcolor}
\usepackage{colortbl}
\usepackage{amsmath,amsfonts}
\usepackage{graphicx}
\usepackage{siunitx}
\usepackage{xspace}
\usepackage{multirow}
\usepackage{enumitem}
\usepackage{makecell}
\usepackage{soul}
\usepackage[ruled]{algorithm2e}
\usepackage[frozencache,cachedir=minted-cache]{minted}
\usepackage[most]{tcolorbox}
\usepackage[
    caption=false,
    font=footnotesize,
    labelfont=footnotesize,
    textfont=footnotesize,
    justification=centering
]{subfig}
\usepackage{hyperref}
\usepackage[nameinlink]{cleveref}
\usepackage{booktabs}
\usepackage{orcidlink}

\hypersetup{
    colorlinks=true,
    linkcolor=blue,
    citecolor=blue,
    filecolor=magenta,
    urlcolor=blue
}

\newsavebox{\unconstrainedtestbox}
\newsavebox{\constrainedtestbox}
\newsavebox{\cbdltestbox}

\definecolor{lightred}{RGB}{255,230,230}  
\newtcolorbox{greenhighlight}{
  colback=green!10,
  colframe=green!10,
  boxrule=0pt,
  sharp corners,
  breakable,
  left=4pt,
  right=4pt,
  top=4pt,
  bottom=4pt
}

\newcommand{\fp}{F$\rightarrow$P\xspace}
\newcommand{\ourtool}{CBDL\xspace}
\newcommand{\llama}{Llama3.3-70B\xspace}
\newcommand{\gptoss}{GPT-OSS-120B\xspace}
\newcommand{\gptcommercial}{GPT-5.2\xspace}

\def\BibTeX{{\rm B\kern-.05em{\sc i\kern-.025em b}\kern-.08em
    T\kern-.1667em\lower.7ex\hbox{E}\kern-.125emX}}
\begin{document}

\title{No More Import Errors: Constrained Decoding for Guaranteed Valid Imports in LLM-Generated Tests}

\author{
    \IEEEauthorblockN{
        Konstantinos Kitsios{\hypersetup{pdfborder={0 0 0}}\orcidlink{0009-0004-1129-3993}}\textsuperscript{1},
        Matteo Biagiola{\hypersetup{pdfborder={0 0 0}}\orcidlink{0000-0002-7825-3409}}\textsuperscript{2,3},
        Guido Salvaneschi{\hypersetup{pdfborder={0 0 0}}\orcidlink{0000-0002-9324-8894}}\textsuperscript{2},
        Alberto Bacchelli{\hypersetup{pdfborder={0 0 0}}\orcidlink{0000-0003-0193-6823}}\textsuperscript{1}
    }\\
    \IEEEauthorblockA{
        \textsuperscript{1}\textit{University of Zurich}, Zurich, Switzerland\\
        konstantinos.kitsios@uzh.ch, bacchelli@ifi.uzh.ch
    }\\
    \IEEEauthorblockA{
        \textsuperscript{2}\textit{University of St.~Gallen}, St.~Gallen, Switzerland\\
        \{guido.salvaneschi, matteo.biagiola\}@unisg.ch
    }\\
    \IEEEauthorblockA{
        \textsuperscript{3}\textit{Università della Svizzera italiana}, Lugano, Switzerland\\
        matteo.biagiola@usi.ch
    }
}

\maketitle
\thispagestyle{plain}
\pagestyle{plain}

\begin{abstract}
One of the most common failure reasons of LLM-based test generation pipelines is \textit{import errors}, e.g., the LLM attempts to import a class that either does not exist or lives in a different package.
Grammar constrained decoding can modify the probabilities of the LLM's final layer during token generation to \textit{guarantee} valid imports.
However, we find that it can lead to degenerate solutions, i.e., repetitive valid imports (\emph{import streaks}) that exhaust the token budget before a test is generated.
We propose Constrained Beam Decoding with Lookup (\ourtool), a novel method that guarantees import validity while addressing import streaks by avoiding premature commitment to tokens that lead to repetitive imports.
Because it operates at the decoding stage, \ourtool can be seamlessly integrated into existing LLM-based test generation pipelines.

We implement \ourtool for Python and Java and empirically evaluate it on two test generation tasks, showing that generated tests have 100\% valid imports, outperforming both static (RAG and LIBRO) and dynamic (Agentic Repair) import-validity approaches that leave 45\%, 10\%, and 43\%, of import errors unresolved.
\ourtool converts on average 14\% of initial import errors into functionally correct tests, increasing overall test correctness by 1--4 percentage points over unconstrained generation.

Even when import validity does not directly lead to a correct test, it allows agentic repair pipelines to allocate the repair budget to remaining errors, increasing correct tests by 2--7 percentage points for Python and 1--6 for Java. 
Finally, we show that a hybrid pipeline, where a small local model equipped with \ourtool generates imports while the frontier \gptcommercial model generates the test body, outperforms generating both with \gptcommercial alone by 1--3 percentage points.

Overall, \ourtool removes a major bottleneck in LLM-based test generation: by guaranteeing import validity with only a small overhead (16--21 seconds per test), it improves test correctness and enables cost-effective hybrid generation.
\end{abstract}


\section{Introduction}\label{sec:introduction}

Large language model (LLM)-based test generation pipelines generate tests by feeding the code-under-test, related context, and a task-specific prompt to an LLM (or LLM-based agent), which generates a test consisting of import statements, a test prefix, and assertions. 
Then, the test is executed to check if it satisfies the target task, e.g., by failing in the buggy and passing in the patched code (\fp), in which case the test is returned as output, otherwise it is discarded or refined in an agentic repair loop.

Such pipelines are of increasing interest both in research and practice~\cite{llms4testing_survey}.
For example, Google generates issue-reproducing tests by feeding an issue description to an LLM pipeline and checking whether the generated test fails in the presence of the issue and passes in its absence, in which case the test is considered an issue-reproducing test, or an \fp test~\cite{google}.
Similarly, Mozilla deployed a tool that is triggered on pull request (PR) creation and feeds the linked issue and patch to an LLM-based pipeline to generate an \fp test that guards against future regressions~\cite{blast}.
Finally, Meta uses an LLM pipeline to generate tests that pass in the current version of the code and improve the coverage of the existing test suite~\cite{meta}.



\begin{figure*}
\centering
\subfloat[Task 1: \fp test generation from issue description and patch.]{%
    \includegraphics[width=0.49\textwidth]{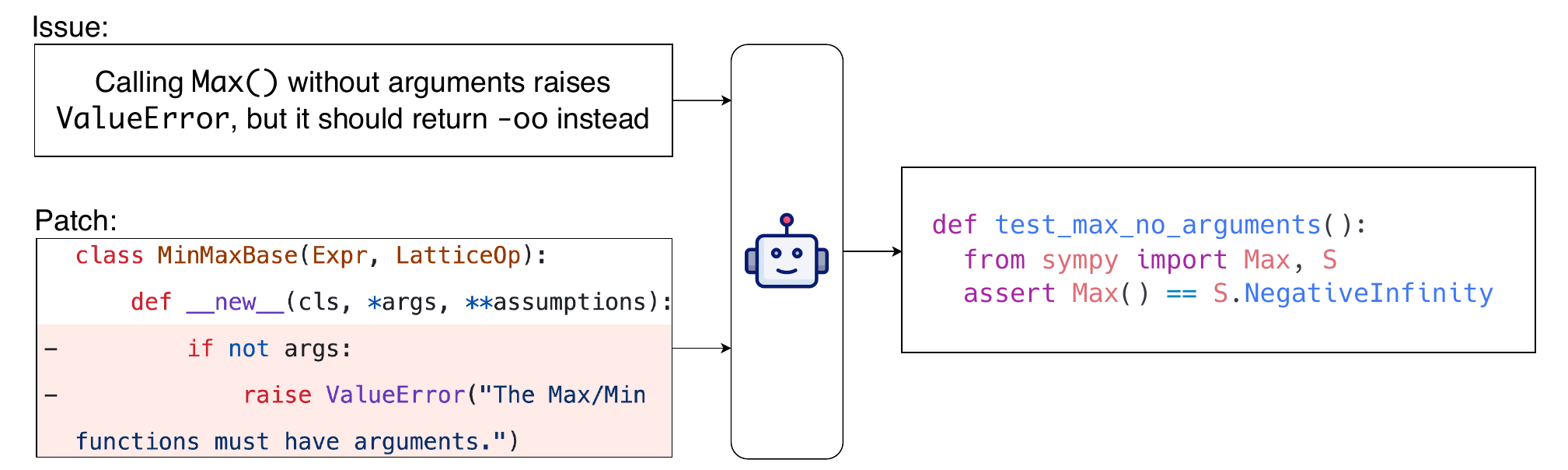}%
}
\hfill
\subfloat[Task 2: Regression test generation with implementation as oracle.]{%
    \includegraphics[width=0.49\textwidth]{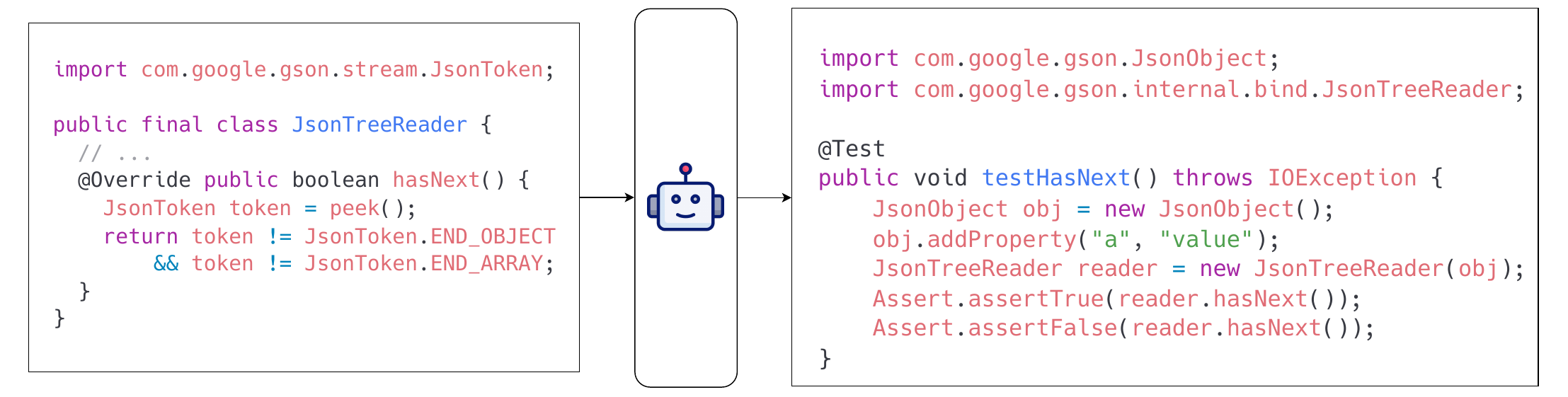}%
}
\caption{The two test generation tasks used for evaluation.}
\label{fig:tasks}
\end{figure*}

However, such pipelines often fail to generate an appropriate test.
Literature~\cite{understanding_agents, understanding_agents2, import_errors_important, package_hallucination, lops6843083benjitest,lops2024agonetest,shah2026characterizing} consistently ranks \textit{import errors} as one of the most prominent causes for this failure.
An \textit{import error} occurs when the LLM-generated test imports an entity (e.g., class, module, package) that does not exist in the runtime environment. 
For example, running \texttt{import com.tools.checkstyle.utils.CommonUtil} yields an import error if the \texttt{CommonUtil} class does not exist or resides in a different package.
Chen et al.~\cite{understanding_agents2} analyzed eight state-of-the-art LLM-based agents and found that import errors were the leading cause of failure, accounting for \num{1515} (23\%) errors across all agents and dataset instances, followed by type errors at \num{992} (15\%).
Our results corroborate these findings: even the state-of-the-art \gptcommercial produces an import error in 18/103 Java test generation tasks.

Motivated by the prevalence of this problem, we develop an approach to \textit{guarantee} import validity for existing LLM-based test generators.
We first observe that all the aforementioned pipelines have access to the execution environment, which allows for valid imports to be statically extracted.
We find that the number of valid import statements in popular Python and Java projects is on the order of \num{100000}, making it impractical to include them directly in the LLM prompt.
Indeed, such long prompts would result in ``lost-in-the-middle'' effects~\cite{liu2024lost} that degrade performance without offering \textit{guarantees} that the LLM would follow instructions~\cite{llm_instruction_following}.

To guarantee import validity \textit{efficiently} we formulate this problem as a constraining problem: generated import statements must belong to a set extracted from the execution environment.
However, we find that enforcing this constraint through greedy grammar constrained decoding~\cite{geng-etal-2023-grammar} leads to \emph{import streaks}, i.e., degenerate solutions where the model repeatedly generates valid imports without proceeding to the test body, aligned with prior findings on low quality output of grammar constrained decoding~\cite{grammar_alinged_decoding,schall2025hidden}.
Our analysis suggests that import streaks are caused by prematurely committing to a locally optimal import, motivating Constrained Beam Decoding with Lookup (\ourtool). 
By retaining alternative continuations, beam search avoids local optima, substantially reducing import streaks. 
A complementary lookup mechanism preserves generated class names when they exist under a different package and activates constrained decoding only when lookup cannot resolve the import. 
Together, these components address both import validity and the repetitive outputs introduced by enforcing it.

We implement \ourtool for Java and Python and empirically evaluate how it improves two existing test generation methods~\cite{blast,chattester} across two test generation tasks, namely \fp test generation and regression test generation, using three LLMs.
We find that \ourtool \textit{guarantees} that all imports are valid, i.e., that the imported entity exists in the runtime environment.
In contrast, other import validity approaches, like retrieval-augmented generation (RAG), LIBRO's dependency resolution~\cite{libro}, and Agentic Repair~\cite{three_magic_number}, leave  45\%, 10\%, and 43\% of import errors unresolved, respectively.
\ourtool also converts 14\% of the import errors into functionally correct tests on average, leading to an overall increase in functional correctness of 1--4 percentage points compared to unconstrained generation.
Even when the tests are not functionally correct, the absence of import errors helps subsequent repair attempts focus on the remaining errors, leading to 1--7 percentage points increase in functional correctness over unconstrained generation.

Furthermore, delegating import generation to a local \ourtool-equipped model while generating the test body with the frontier \gptcommercial yields more correct tests than generating both with \gptcommercial.
Finally, we show that \ourtool adds an average overhead of 16--21 seconds per test as it enforces validity at generation time instead of relying on costly repair loops.
Overall, our work eliminates a major failure reason in LLM-based test generation through a novel, guarantee-based approach, leading to the following contributions:

\begin{itemize}
    \item A formulation of import validity as constraining problem.
    \item \ourtool, a novel approach that guarantees import validity, mitigates import streaks via beam search, and can be integrated with existing LLM-based test generators.
    \item Empirical evidence that \ourtool outperforms five static and dynamic import-validity approaches in terms of import validity and test correctness.
    \item Empirical evidence that letting a small local model with \ourtool write the imports and a remote frontier LLM write the test body outperforms fully remote generation.
\end{itemize}

\begin{algorithm}
\scriptsize
\LinesNumbered
\caption{Unconstrained LLM Generation.}
\label{alg:autoregressive_decoding}
\KwIn{Language model $f_\theta$, prompt $c$}
\KwOut{Generated sequence $y$}
$y \leftarrow [\,]$\;
\While{True}{
    compute next-token distribution $P_t = f_\theta(c, y)$\;
    select next token $y_t = \arg\max(P_t)$\; 
    append $y_t$ to $y$\;
    \If{$y_t$ is \texttt{EOS} or limit reached}{
        \textbf{break}\;
    }
}
\Return{$y$}
\end{algorithm}

\section{Background}\label{sec:background}

\subsection{How Do LLMs Generate Code?}\label{ssec:background_llms}
LLMs operate over a predefined vocabulary of tokens $\mathcal{V}$, where a token is a sequence of $\geq 1$ characters.
Given an input prompt~$c$ (e.g., natural language instructions, source code, or both), the model generates a sequence of tokens one at a time, following the process described in \autoref{alg:autoregressive_decoding}. 
At each step, the model takes as input the prompt concatenated with the tokens generated so far and computes a probability distribution $P_t$ over all tokens in $\mathcal{V}$ (Line~3). 
A decoding strategy then selects the next token $y_t$ from this distribution, e.g., by choosing the most probable token (Line~4).
Finally, the selected token $y_t$ is appended to the sequence (Line~5) and the process repeats until a termination condition is reached, such as generating an end-of-sequence token or reaching a maximum output length (Lines~6--7).

\subsection{LLM-Based Test Generation}\label{ssec:background_testgen}
We introduce the two test generation tasks used in our evaluation, which are schematically illustrated in \Cref{fig:tasks} (a) and (b) respectively.
In the task of \textbf{\fp test generation (Task~1)}~\cite{blast,otter,tddbench,swtbench,libro}, the inputs are a bug description and optionally the patch that fixes the bug.
The generated test must expose the reported bug by failing before the patch is applied and passing afterward. It then serves as a regression test against reintroducing the same bug.
We provide both the issue description and its patch to the LLM~\cite{blast,swtbench}, demonstrating that invalid imports persist even when the model receives the corresponding fix.

In the task of \textbf{regression test generation (Task~2)}~\cite{chattester,chatunittest,codamosa}, the input is a method/function, and the output is a test that asserts the current behavior of the method/function.
Thus, the current behavior is considered the oracle, and the goal is to ensure that future modifications do not break it.
Existing work in these tasks is discussed in \Cref{sec:related_work}.

\section{Problem Statement}

Import errors are a practical obstacle in AI-based software development. 
In the 2026 Stack Overflow Developer Survey, 45.1\% of 13,648 respondents selected ``Hallucinated package/config'' among the mistakes that had cost them the most time, making it the fourth most common category~\cite{stackoverflow}.
Although this category encompasses both package and configuration
hallucinations, it highlights the practical burden of generated code
that relies on incorrectly identified dependencies.

Agents do not eliminate this problem: Chen et al.~\cite{understanding_agents2} analyzed eight agents and identified import errors as the most frequent error category in agent trajectories, accounting for 1,515 (23\%) of the agents' errors, followed by type errors at 992 (15\%).
Although agents can use execution feedback to revise their solutions, import errors persisted into the final submitted solution across \textit{all} eight agents.
Other agents or LLM-based pipelines in the literature also report import errors as a leading failure mode~\cite{understanding_agents, understanding_agents2, import_errors_important, package_hallucination, lops6843083benjitest,lops2024agonetest,shah2026characterizing}.

Our evaluation provides complementary evidence in the specific
setting of test generation (\Cref{ssec:rq4_results}). 
GPT-5.2, a frontier commercial model at the time of our experiments, generated import errors in 18 of the 103 Java F$\rightarrow$P test-generation tasks (17.5\%).
This demonstrates that strong code generation capabilities alone do not ensure import validity or alignment with the imports available in a target project. 

Together, the developer survey, the agent study,
and the GPT-5.2 experiments suggest that import validity is a major bottleneck in LLM-based code generation, calling for specific solutions rather than relying on models or agents to infer valid imports or repair violations.

\section{Constrained Beam Decoding with Lookup}

Our work, implemented in \ourtool, addresses the problem of the validity of generated imports by \textit{enforcing} such validity during generation, allowing the remaining generation and repair budget to focus on the test body and its assertions.\\

In this section, we first motivate \ourtool with an example (\Cref{ssec:meth_motivating_example}).
Then, we present the stages of \ourtool, i.e., how it extracts valid import statements from the execution environment (\Cref{ssec:meth_extracting_valid_grammars}), constrains the LLM output to adhere to these import statements (\Cref{ssec:meth_vanilla}), prevents distorted output (\Cref{ssec:method_beam}), and uses a lookup mechanism to improve effectiveness and efficiency (\Cref{ssec:method_lookup}).
\ourtool requires an open-weight LLM and can be integrated with existing state-of-the-art test generators that use such models without modifications.

\begin{figure}
  \centering
  \includegraphics[width=0.48\textwidth]{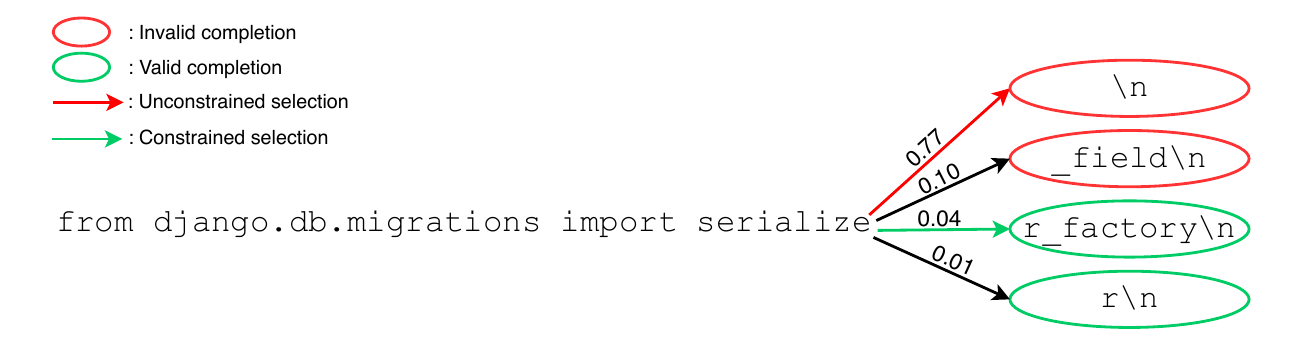}
  \caption{Example of constrained decoding selecting a valid completion. While the unconstrained model selects the most probable token that leads to an invalid import, constrained decoding restricts the choices to valid completions and selects the most probable among them.}
  \label{fig:motivation}
\end{figure}

\subsection{Motivating Example}\label{ssec:meth_motivating_example}
When prompted to generate an \fp test for the \texttt{django-12125} instance of the SWT-Bench~\cite{swtbench} dataset, \gptoss starts generating the import statement \texttt{from django.db.migrations import serialize}. 
The top four completions for the next token are shown in \Cref{fig:motivation}.
The unconstrained generation selects the newline character as the next token because it has the highest probability, 0.77.
However, this completion is invalid, as there is no \texttt{serialize} class in the \texttt{django.db.migrations} module, so the generated test yields an import error.

\ourtool avoids this error because it enforces only valid imports (green in \Cref{fig:motivation}), i.e., completions \#3 and \#4.
Out of these, \ourtool selects \#3 because, according to the learned distribution, it is four times more likely than \#4.
Thus, \ourtool guarantees that invalid imports are never selected, while using the learned distribution to select the most likely valid import. 

\subsection{Extracting Valid Import Statements}\label{ssec:meth_extracting_valid_grammars}
The first step of \ourtool is to extract a list of all valid import statements for a given execution environment.
The extraction is language-specific and \ourtool supports two popular programming languages, i.e., Java and Python.
For this step \ourtool needs static access to the execution environment, which is provided by common test generation benchmarks~\cite{swtbench,defects4j} and used by existing test generators~\cite{blast, meta, google} \ourtool is integrated with.

\subsubsection{\textbf{Java}} The official Java Language Specification (JLS)~\cite{jls17_imports} defines four forms of import statements, shown in the upper half of~\Cref{tab:import_types}, along with their syntax and description.
To extract all valid statements of the form \#1, it suffices to extract all types from all packages in the Java classpath of each project.
To do this, \ourtool compiles the project and then uses the ClassGraph tool~\cite{classgraph} to statically scan the compiled project and enumerate all available classes, including the project's compiled classes, their dependencies, and standard modules.
\ourtool then filters out private and anonymous
classes that cannot be imported.
The resulting set of fully qualified class names corresponds to all valid import statements of form \#1.

From this set, obtaining valid import statements for forms \#2 and \#4 is straightforward: 
if \texttt{import X.Y;} is valid, then \texttt{import X.*;} and \texttt{import static X.Y.*;} are valid.
Regarding form \#3, we could not find a reliable way to obtain all \textit{public} static members of a class due to inheritance rules that create cross-class dependencies.
However, form \#3 is subsumed by form \#4. 
Instead of writing \texttt{import static java.lang.Math.PI;}, one can write \texttt{import static java.lang.Math.*;}, which also imports the \texttt{PI} constant of the \texttt{Math} class.
Thus, \ourtool supports import forms \#1, \#2, and \#4.
The resulting corpus of import statements forms the grammar used by \ourtool for constrained decoding.

\begin{table}
\scriptsize
\centering
\caption{Import statement syntax for Java and Python.}

\setlength{\tabcolsep}{1.2pt}

\begin{tabular}{lllp{2.3cm}}
\textbf{Lang.} & \textbf{Import Form} & \textbf{Syntax} & \textbf{Description} \\
\toprule
Java & 1-SingleTypeImport     & \texttt{import X.Y;}          & X is a package, Y is a type (e.g., class) of X. \\
 \rowcolor{gray!20}
Java & 2-TypeImportOnDemand   & \texttt{import X.*;}        & X is a package. \\
Java & 3-SingleStaticImport   & \texttt{import static X.Y.Z;} & X.Y is a type (e.g., class), Z is a static member of X.Y. \\
 \rowcolor{gray!20}
Java & 4-StaticImportOnDemand & \texttt{import static X.Y.*;} & X.Y is a type. \\
\midrule
Python & 1-Module Import           & \texttt{import X.Y}        & X.Y is a module. \\
 \rowcolor{gray!20}
Python & 2-Single Attribute Import & \texttt{from X.Y import Z} & X.Y is a module, Z is an attribute (e.g., class) of X.Y. \\
Python & 3-All Attributes Import    & \texttt{from X.Y import *} & X.Y is a module. \\

\bottomrule
\end{tabular}
\label{tab:import_types}
\end{table}

\subsubsection{\textbf{Python}} From the Python Language Reference~\cite{python_import_stmt} we extract the three forms shown in the bottom half of~\Cref{tab:import_types}.
In the absence of an out-of-the-box tool like ClassGraph for Python, we use \texttt{sys.builtin\_module\_names} method to obtain built-in modules (e.g., \texttt{math}, \texttt{sys}) and \texttt{pkgutil.iter\_modules()} to enumerate installed and local modules (e.g., \texttt{numpy}). 
This yields the full path of top-level modules such as \texttt{numpy}, but not their submodules (e.g., \texttt{numpy.linalg.eig}). 
To discover these, we recursively call \texttt{pkgutil.iter\_modules()} from the subdirectories of the discovered modules, thereby enumerating all submodules and obtaining the valid imports of forms \#1 and \#3.

Finally, to obtain the attributes (e.g., classes or functions) of each discovered module, we statically parse the module's \texttt{\_\_init\_\_.py} file using Python's \texttt{ast} module.
If \texttt{\_\_init\_\_.py} defines the variable \texttt{\_\_all\_\_}, we treat it as the list of exported attributes, otherwise we collect all public top-level definitions.
This yields the list of valid imports of form \#2, which together with the list of forms \#1 and \#3 define the valid import grammar for Python.

\begin{algorithm}
\scriptsize
\LinesNumbered
\caption{Constrained Decoding}
\label{alg:constrained_decoding}
\KwIn{Language model $f_\theta$, prompt $c$ ending with an import keyword, list of available imports $A$}
\KwOut{Generated import statement $y$}
$y \gets [\,]$\;
\While{True}{
    $P_{\textit{unc}} \gets f_\theta(c, y)$\;
    $M \gets [0]^{|\mathcal{V}|}$\;
    \For{each token $t_i \in \mathcal{V}$}{
        \If{$t_i$ is a prefix of some import in $A$}{
            $M_i \gets 1$\;
        }
    }
    $P_{\textit{con}} \propto P_{\textit{unc}} \odot M$\;
    $y_t \gets \arg\max(P_{\textit{con}})$\;
    append $y_t$ to $y$\;
    $A \gets \{a.\texttt{replace}(y_t, \varepsilon) \mid a \in A \mid a \text{ starts with } y_t\}$\;
    \If{$y_t$ is newline}{
        \textbf{break}\;
    }
}

\Return{$y$}
\end{algorithm}

\subsection{Constrained Decoding}\label{ssec:meth_vanilla}

\begin{figure*}[t]
\centering

\begin{lrbox}{\unconstrainedtestbox}
\begin{minipage}[t]{0.3\textwidth}
\vspace{0pt}
\begin{minted}[fontsize=\tiny,highlightlines={2},highlightcolor=lightred]{python}
def test_unconstrained():
    import numpy as np # Import error
    import sympy as sp
    from sympy import Matrix
    from sympy.polys.matrices import normal_form

    matrix = Matrix(np.array([[5, 8, 12], [0, 0, 1]]))
    hnf = normal_form(matrix.T).T
    # ...
\end{minted}
\end{minipage}
\end{lrbox}

\begin{lrbox}{\constrainedtestbox}
\begin{minipage}[t]{0.3\textwidth}
\vspace{0pt}
\begin{minted}[fontsize=\tiny]{python}
def test_constrained():
    import sympy as sp
    import sympy.polys.matrices as spm
    import sympy.polys.domains as spdom
    import sympy.polys.matrices.domainscalar
    import sympy.polys.matrices.domainscalar
    import sympy.polys.matrices.domainscalar
    import sympy.polys.matrices.domainscalar
    # ...
\end{minted}
\end{minipage}
\end{lrbox}

\begin{lrbox}{\cbdltestbox}
\begin{minipage}[t]{0.3\textwidth}
\vspace{0pt}
\begin{minted}[fontsize=\tiny]{python}
def test_cbdl():
    import sympy as sp
    from sympy import Matrix
    from sympy.polys.matrices import normal_form

    A = Matrix([[5, 8, 12], [0, 0, 1]])
    hnf = normal_form(A.T)
    # ...

\end{minted}
\end{minipage}
\end{lrbox}

\subfloat[Unconstrained generation leads to an import error
(highlighted in red).]{\usebox{\unconstrainedtestbox}}
\hfill
\subfloat[Constrained decoding avoids the import error but
leads to an import streak.]{\usebox{\constrainedtestbox}}
\hfill
\subfloat[\ourtool avoids the import error and does not lead
to an import streak.]{\usebox{\cbdltestbox}}

\caption{Comparison of tests generated by (a) unconstrained
generation, (b) constrained generation, and (c) \ourtool.}
\label{fig:beam_tests}
\end{figure*}

Given the corpus of valid import statements extracted from the execution environment (\autoref{ssec:meth_extracting_valid_grammars}), the goal of \ourtool is to ensure that the language model generates only imports that belong to this set. 
In particular, \ourtool activates constrained decoding when the model generates a token sequence that marks the beginning of an import statement. 
According to \Cref{tab:import_types}, the trigger sequences are \{`\texttt{\textbackslash nimport }'\} for Java and \{`\texttt{\textbackslash nimport }', `\texttt{\textbackslash nfrom }'\} for Python. 
The trailing white space ensures that constraining is not accidentally triggered for identifiers such as \texttt{\textbackslash nimportantVariable=10}. 

Once the model generates a trigger sequence, \ourtool switches to constrained mode, and uses \autoref{alg:constrained_decoding} to generate the remainder of the import statement.
The inputs to the constrained decoding algorithm are the language model $f_{\theta}$, the prompt $c$ ending with a trigger sequence, and the corpus of available import statements $A$. 
The output is an import statement $y \in A$, which is initialized to the empty sequence (Line~1). 
Then, the unconstrained next-token probability $P_{\textit{unc}} \in \mathbb{R}^{|\mathcal{V}|}$ is computed (Line~3).
To set the elements of $P_{\textit{unc}}$ that correspond to invalid imports to 0, \ourtool constructs a binary mask $M \in \{0, 1\}^{|\mathcal{V}|}$ where for each token $t_i \in \mathcal{V}$ the mask value $M_i$ is $1$ if $t_i$ is a prefix of at least one import statement in $A$ and $0$ otherwise (Lines~4--9).

The constrained probability distribution $P_{\textit{con}}$ is then obtained by applying the mask to the unconstrained distribution (Line~10). 
The next token $y_t$ is selected from $P_{\textit{con}}$ (Line~11) and appended to the generated sequence (Line~12).
Finally, the corpus of valid imports $A$ is updated to contain all the valid continuations of the import statement generated so far (Line~13).
Constrained decoding continues until a newline token is generated in Python or a semicolon in Java (Line~14), which marks the end of the import statement and \ourtool resumes unconstrained generation. 
Then, the model can either start another import statement, in which case \ourtool re-activates constrained decoding, or move on to the test body.

\subsection{Beam Constrained Decoding}\label{ssec:method_beam}

\begin{figure}
  \centering
  \includegraphics[width=0.5\textwidth]{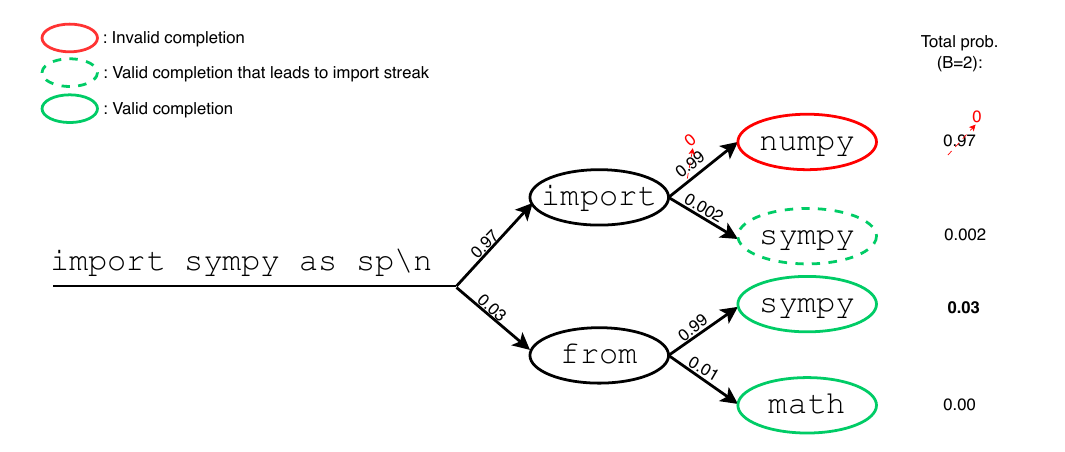}
  \caption{Example of how beam constrained decoding avoids import streaks. Instead of greedily committing to the \texttt{import} completion that would lead to an import streak, it explores both \texttt{import} and \texttt{from} completions and prefers the \texttt{from} because it leads to higher overall probability when expanded.}
  \label{fig:beam}
\end{figure}

The constrained decoding algorithm described in \Cref{ssec:meth_vanilla} modifies the model's learned probability distribution $P_{\textit{unc}}$, which can lead to unusual, low-quality outputs~\cite{grammar_alinged_decoding,schall2025hidden}.
We observed this in our experiments, where the model kept generating valid imports but without ever starting to generate the test body.
We identified the root cause behind this behavior, which we present through an example below, before proposing a solution.

\subsubsection{Motivating Example}\label{sssec:beam_example}

When tasked with generating an \fp test for the \texttt{sympy-23413} instance of SWT-Bench, \llama generates the test shown in \Cref{fig:beam_tests} (a), which yields an import error because \texttt{numpy} is not a valid test dependency in the \texttt{sympy} repository.\footnote{\href{https://github.com/sympy/sympy/blob/master/requirements-dev.txt}{https://github.com/sympy/sympy/blob/master/requirements-dev.txt}}
Using the constrained decoding described in~\autoref{alg:constrained_decoding}, the model generates the test shown in \Cref{fig:beam_tests} (b), where the import error is avoided, but the generation never proceeds to the test body.
By examining $P_{\textit{unc}}$ after each \texttt{import} keyword, we find that the model begins each line with the goal of importing \texttt{numpy}.
Since constraining starts only after the \texttt{import} keyword is generated, the model cannot know that \texttt{import numpy} is invalid, so it keeps generating the \texttt{import} keyword instead of generating \texttt{from sympy import Matrix}, which is the next statement in \Cref{fig:beam_tests} (a).

\subsubsection{Beam Search}\label{sssec:beam_solution}

We call the phenomenon in \Cref{fig:beam_tests} (b) an \textit{import streak}.
To deal with the problem of import streaks, we need a way of ``hinting'' to the model that \texttt{import numpy} is invalid even before generating the \texttt{import} keyword.
To this end, we use \textit{Beam Constrained Decoding}, by combining constrained decoding with beam search~\cite{beam, anderson2017guided}.
Beam search keeps track of $B>1$ completions of the initial prompt instead of greedily committing to the most probable next token.
To illustrate how beam search deals with import streaks, we use the example in \Cref{fig:beam_tests} (b). 
Let us assume that the model has already generated the first import and is about to generate the next one.
The two most probable completions are shown in \Cref{fig:beam}, namely \texttt{import} with probability 0.97 and \texttt{from} with probability 0.03.
Instead of committing to the most probable completion, beam search with $B=2$ expands both completions using the two most probable tokens, calculates the total probability by multiplying the token probabilities (right-most in \Cref{fig:beam}), and retains the top two candidates.
This way, the model can see that the continuation \texttt{import numpy} is not valid before committing to the keyword \texttt{import}, therefore preferring the \texttt{from sympy} completion and leading to the generated test in \Cref{fig:beam_tests} (c).
In the evaluation we use \textit{B=2} because in our ablation study it yielded the best performance.

\ourtool enables beam search only when generating imports, i.e., each beam terminates after the first non-import line. 
After all beams terminate, the highest-probability beam is selected and the generation continues unconstrained to the test body.

\subsection{Lookup Mechanism}\label{ssec:method_lookup}

The final component of \ourtool is a lookup mechanism that improves the effectiveness against a certain type of import errors while also reducing the overhead of constrained decoding.

During initial experiments, we observed that \ourtool occasionally struggled with imports where the class name is correct but the package prefix is incorrect.
For instance, when tasked to generate an \fp test for the \texttt{rocketmq-27} instance of the GHRB dataset~\cite{ghrb}, \gptoss needs to import the class \texttt{ConsumeInitMode}.
Under unconstrained decoding, it generates the statement \texttt{import org.apache.\textbf{consumer}.ConsumeInitMode}, where the package \texttt{org.apache.\textbf{consumer}} is valid but it does not contain the class \texttt{ConsumeInitMode}, which actually resides in the package \texttt{org.apache.\textbf{constant}}.
Hence, the model ends up importing an invalid class from the \texttt{consumer} package, resulting in an incorrect test.

To address this limitation, \ourtool introduces a lookup mechanism that works as follows.
\ourtool does not switch to constrained decoding immediately after the first trigger keyword, rather it starts generating imports in unconstrained mode until an invalid import is generated.
When this happens, \ourtool performs a lookup to determine whether the class (suffix) exists in a different package from the extracted list of valid imports.
If such a match is found, \ourtool replaces the incorrect package prefix with the correct one and continues the generation.
Otherwise, the generated import is discarded and \ourtool switches to beam constrained decoding (\Cref{ssec:method_beam}) to generate the remaining imports.
If more than one match is found, the package with the highest character similarity with the initial package is selected.
This hybrid strategy also reduces the overhead of \ourtool since it triggers constrained decoding only when unconstrained generation would lead to an import error.

\subsection{Limitations}\label{ssec:meth_limitations}
\ourtool guarantees that all the generated import statements are \textit{valid}, which does not necessarily mean that they are \textit{appropriate} to solve the given task.
For example, if the unconstrained probability of an import statement required for an \fp test is zero, constrained decoding cannot enforce the generation of such import.
Another limitation concerns the import extraction stage, where our approach captures the majority of statically defined imports, but it may miss cases such as C-extension modules in Python or dynamically defined exports (e.g., through string operations) in \texttt{\_\_init\_\_.py}.
We leave more precise import extraction, including approaches based on language servers~\cite{lsp}, to future work.

\section{Implementation}\label{sec:impl}
We implemented \ourtool on top of the \texttt{llama.cpp} library that is widely used to host models locally.
Specifically, we modified the completion API to specify tokens that are allowed at each generation step.
On the server side, we then set the logits of all the tokens that are not allowed to $-\infty$, 
which effectively turns their probability to zero after the softmax layer.
Delegating constraining operations to server-side C++ code keeps the performance overhead low.
We implemented the client side in Python, which computes the allowed tokens at each step given a partially generated import and the list of valid imports.

We run the \texttt{llama.cpp} server on a Mac Studio with an Apple M3 Chip and \num{256} GB of unified memory for the Python experiments and on an HP Z1 Mini with an AMD Ryzen AI Max+ Pro 395 and 128 GB of unified memory for the Java experiments.
We configure both servers with a context window of 24K tokens.

\section{Evaluation}\label{sec:eval}

To evaluate the benefits of \ourtool, we formulated the following research questions:

\begin{itemize}
    \item \textbf{RQ1.} To what extent can \ourtool generate tests without import errors while avoiding import streaks?

    \item \textbf{RQ2.} What is the effect of guaranteed import validity on functional correctness?

    \item \textbf{RQ3.} How do the components and hyperparameters of \ourtool affect its performance?

    \item \textbf{RQ4.} How does the performance of a state-of-the-art, commercial LLM change if we delegate the import generation to a smaller, local model equipped with \ourtool?

    \item \textbf{RQ5.} What is the runtime overhead of \ourtool?
\end{itemize}

\begin{table*}
    \centering
    \footnotesize
    \setlength{\tabcolsep}{7.2pt}
    \renewcommand{\arraystretch}{0.8}
    \caption{Number of import errors and import streaks for each import-validity approach, across two benchmarks in two languages, two test generation tasks and two models.}
    \begin{tabular}{llrrrrrrrr}
        & & \multicolumn{4}{c}{\textbf{\# Import Errors ($\downarrow=$ better)}} & \multicolumn{4}{c}{\textbf{\# Import Streaks ($\downarrow=$ better)}} \\
        \cmidrule(lr){3-6} \cmidrule(lr){7-10}
        & & \multicolumn{2}{c}{\textbf{\shortstack{\fp Test Generation\\(BLAST)}}} & \multicolumn{2}{c}{\textbf{\shortstack{Regression Test\\Generation (ChatTester)}}} & \multicolumn{2}{c}{\textbf{\shortstack{\fp Test Generation\\(BLAST)}}} & \multicolumn{2}{c}{\textbf{\shortstack{Regression Test\\Generation (ChatTester)}}}\\
        \cmidrule(lr){3-4} \cmidrule(lr){5-6} \cmidrule(lr){7-8} \cmidrule(lr){9-10}
        \textbf{Language} & \textbf{\shortstack{Import Validity\\Approach}} & Llama & GPT-OSS & Llama & GPT-OSS & Llama & GPT-OSS & Llama & GPT-OSS \\
        \midrule
\multirow{6}{*}{\shortstack{Java \\ (N=103)}} & Unconstrained & 36 & 22 & 24 & 11 & 0 & 0 & 0 & 1 \\
 & Libro DR & 4 & 1 & 0 & 1 & 0 & 0 & 0 & 1 \\
 & RAG & 11 & 17 & 9 & 9 & 0 & 0 & 0 & 0 \\
 & Agentic Repair & 28 & 13 & 13 & 7 & 0 & 0 & 0 & 0 \\
 & Constrained Decoding & 0 & 0 & 0 & 0 & 12 & 16 & 7 & 9 \\
 & CBDL & 0 & 0 & 0 & 0 & 2 & 1 & 4 & 0 \\
        
        \midrule
\multirow{6}{*}{\shortstack{Python \\ (N=323)}} & Unconstrained & 29 & 26 & 20 & 38 & 0 & 0 & 0 & 0 \\
 & Libro DR & 7 & 1 & 6 & 0 & 0 & 0 & 0 & 0 \\
 & RAG & 15 & 9 & 9 & 13 & 0 & 0 & 0 & 0 \\
 & Agentic Repair & 16 & 5 & 5 & 2 & 0 & 0 & 0 & 0 \\
 & Constrained Decoding & 0 & 0 & 0 & 0 & 3 & 9 & 6 & 16 \\
 & CBDL & 0 & 0 & 0 & 0 & 0 & 1 & 1 & 4 \\
        \bottomrule
        \multicolumn{10}{l}{Llama: \llama, GPT-OSS: \gptoss}
    \end{tabular}
    \label{tab:res_rq1}
\end{table*}

\subsection{Experimental Setup}

\subsubsection*{\textbf{Datasets}}
We evaluate \ourtool on the two test generation tasks introduced in \Cref{sec:background}, namely \textit{\fp Test Generation} (Task~1) and \textit{Regression Test Generation} (Task~2), across two programming languages, namely Python and Java.
For \textbf{Task 1 in Java}, we use the GitHub Recent Bugs (GHRB) dataset~\cite{ghrb,libro}, which pairs bugs in 16 popular Java repositories with their fixes and is updated every three months with new bugs, helping reduce the risk of contamination from model training data.
We use commit \texttt{ed4a6ed} as the cutoff, resulting in \num{103} bugs after removing \num{18} due to build failures, \num{14} of which from the \texttt{jackson} project.

We further adapt GHRB for \textbf{Task 2 in Java} as follows.
From each patch, we extract the fixed version of the modified method and treat it as the correct version for which we want to generate passing tests. 
If more than one method is modified, we instruct the LLM to generate a passing test for the first one.
We also extract all global statements of the method's file, as well as the class signature and constructor to use them as context for the LLM.
Note that the global statements include the file's imports, so any import errors generated by the LLM correspond to additional imports required for the testing task.
This process yields \num{103} fixed method implementations, an example of which is shown in \Cref{fig:tasks} (b).

For \textbf{Task 1 in Python}, we use the SWT-Bench-Verified~\cite{tddbench} dataset, which comprises \num{449} GitHub issues from \num{12} popular Python repositories along with the corresponding patch.
We remove the \num{18} instances of the \texttt{astropy} repository because the benchmark does not report the detailed error message for failing tests so we could not know whether a test failure was due to an import error or another error.
We also remove \num{53} instances that yielded errors unrelated to the tests, such as failure to set up the project due to broken dependencies or timeouts after \num{30} minutes, as well as \num{55} instances that require an x86 architecture and could not be executed in our ARM environment, leaving us with a total of \num{323} instances.
Finally, we adapt SWT-Bench-Verified for \textbf{Task 2 in Python} following the same steps as in Java.

\subsubsection*{\textbf{Underlying Test Generation Methods and LLMs}}
CBDL is applied on top of existing LLM-based test generation methods to eliminate their import errors. 
For Task 1, we apply \ourtool on top of BLAST~\cite{blast}, a recent pipeline for generating \fp tests that extracts relevant context through static analysis and analyzes Git history to identify the appropriate file to place the generated test. 
Similarly, for Task 2, we apply \ourtool on top of the carefully designed prompt outline of ChatTester~\cite{chattester}; for our single-query experiments, we disable the test refinement loop of ChatTester and replace it with our agentic repair loop for the multi-query experiments. Since both approaches were designed for Python, we slightly adjust the prompts for Java.

As underlying open-source LLMs we select the non-reasoning model \texttt{
Llama-3.3-70B-Instruct}~\cite{llama3.3} and the reasoning model \texttt{GPT-OSS-120B}~\cite{openai2025gptoss} quantized to 16 bits.
For the combination with commercial LLMs in RQ4, we select \gptcommercial, a frontier reasoning model from OpenAI.
We pin the model version to \texttt{gpt-5.2-2025-12-11} to increase reproducibility and use the official OpenAI API to query it.

\subsubsection*{\textbf{Baselines}}
We compare \ourtool with five \textit{import-validity} baselines.
All baselines use the two test generation backbones (i.e., BLAST~\cite{blast} and ChatTester~\cite{chattester}), addressing import validity either before or after the generation.

The first baseline is the original unconstrained generation (i.e., no import validity check), which is expected to generate import errors.

The second baseline is LIBRO's dependency resolution (LIBRO DR)~\cite{libro}, which takes as input the test body without imports and statically infers the required imports from the repository.
Given a generated test method, LIBRO parses it to extract variable types and referenced class names, constructors, and exceptions, adding resolved imports to the test. 
For Java, we use the authors' publicly released replication code, which we also adapt for Python.

The third baseline is a retrieval-augmented generation (RAG) approach, which we use to test whether including valid imports in the prompt is sufficient to mitigate import errors.
In particular, the RAG baseline retrieves a subset of valid imports given the corpus of valid imports extracted from the execution environment (on the order of \num{100000}s), avoiding context overloading.
To do this, it extracts all non-keyword identifiers from the input patch or code. 
For instance, from the patch in \Cref{fig:tasks} (a) it extracts \texttt{MinMaxBase}, \texttt{Expr}, and \texttt{LatticeOp}. 
Then, it searches the corpus for these classes, and appends them to the prompt if found.
The fourth baseline is grammar constrained decoding~\cite{geng-etal-2023-grammar}, which enforces import validity at generation time but suffers from import streaks.

Finally, our fifth baseline is an agentic repair pipeline that attempts to repair import errors after the generation, which is commonly used in state-of-the-art agentic approaches~\cite{three_magic_number}.
It first calls unconstrained BLAST or ChatTester to generate a test, runs it, and collects the error logs.
If the logs contain an import error, it adds the error details to the original prompt, asking the model to generate a revised test without these errors and repeating until the test is successful or the query budget is exhausted.
We limit the repair loop to three iterations, as prior work showed diminishing gains beyond this~\cite{three_magic_number}.

To evaluate whether eliminating import errors allows the repair budget to address other errors and ultimately produce more correct tests, we also evaluate CBDL followed by repair. 
Starting with a CBDL generation that guarantees a test without import errors, we allow up to two unconstrained repair attempts, matching the number of queries of Agentic Repair.

To compare the import-error-repair capabilities of the above baselines, we run them on the dataset instances where unconstrained generation led to an import error and retain the unconstrained outputs unchanged for the remaining instances.

\subsubsection*{\textbf{Mitigating LLM Non-Determinism}}\label{sssec:non_determinism}
The \llama model comes with a temperature parameter that we set to \num{0} to reduce the non-determinism as done in previous studies~\cite{blast, tddbench, otter}.
However, \gptoss and \gptcommercial do not expose a user-controlled temperature parameter.
Thus, for these models we repeat each test generation instance three times and report the median of each metric.
To further mitigate the effect of non-determinism for \gptoss, we reuse the reasoning trace of the unconstrained generation for all import validity baselines, thereby controlling variation in reasoning when comparing import validity strategies.

\begin{figure}
\centering
\begin{minted}[fontsize=\scriptsize, highlightlines={1-2, 4-5, 7-8},highlightcolor=lightred]{python} 
# (a) Agentic repair 1st attempt -> Import error
from django.db.migrations.serializer import serialize

# (b) Agentic repair 2nd attempt -> Different import error
from django.db.migrations.serializer import ClassSerializer

# (c) Agentic repair 3rd attempt -> Same import error
from django.db.migrations.serializer import ClassSerializer

# (d) CBDL 1st attempt -> valid import led to F2P test
from django.db.migrations.serializer import serializer_factory
\end{minted}
\caption{Example where an agentic repair pipeline fails to fix an import error after three attempts (a--c), while \ourtool generates a valid import in a single attempt (d).}\label{fig:agentic_failure}
\end{figure}

\begin{table*}
    \centering
\footnotesize
\setlength{\tabcolsep}{7pt}
\renewcommand{\arraystretch}{0.81}
    \caption{Number of functionally correct tests ($\uparrow=$ better).}
    \begin{tabular*}{\textwidth}{@{\extracolsep{\fill}}lllrrrr@{}}
        \textbf{Language} & \textbf{\# Attempts} & \textbf{Import Validity Approach} & \multicolumn{2}{c}{\textbf{\shortstack{\fp Test Generation\\(BLAST)}}} & \multicolumn{2}{c}{\textbf{\shortstack{Regression Test Generation\\(ChatTester)}}} \\
        \cmidrule(lr){4-5} \cmidrule(lr){6-7}
        & & & \textbf{Llama} & \textbf{GPT-OSS} & \textbf{Llama} & \textbf{GPT-OSS} \\
        \midrule
\multirow{6}{*}{\shortstack{Java \\ (N=103)}} & \multirow{4}{*}{1} & Unconstrained & 8 & 34 & 10 & 36 \\
 &  & Libro DR & 8 & 35 & \textbf{11} & 36 \\
 &  & RAG & 9 & 36 & 10 & 36 \\
 &  & CBDL & \textbf{10} & \textbf{38} & 10 & \textbf{37} \\
        \cmidrule(lr){2-7}
 & \multirow{2}{*}{Up to 3} & Agentic Repair & \textbf{10} & 37 & \textbf{11} & 37 \\
 &  & CBDL & \textbf{10} & \textbf{40} & \textbf{11} & \textbf{40} \\

        \midrule
        \midrule

\multirow{6}{*}{\shortstack{Python \\ (N=323)}} & \multirow{4}{*}{1} & Unconstrained & 72 & 135 & 86 & 144 \\
 &  & Libro DR & 73 & 137 & 86 & 148 \\
 &  & RAG & \textbf{75} & 140 & 86 & 155 \\
 &  & CBDL & \textbf{75} & \textbf{141} & \textbf{88} & \textbf{157} \\
        \cmidrule(lr){2-7}
 & \multirow{2}{*}{Up to 3} & Agentic Repair & 73 & \textbf{146} & 87 & 163 \\
 &  & CBDL & \textbf{78} & 144 & \textbf{93} & \textbf{167} \\
 \bottomrule
        \multicolumn{7}{l}{Llama: \llama, GPT-OSS: \gptoss}
\end{tabular*}
    \label{tab:res_rq2}
\end{table*}

\subsection{RQ1---Guarantees For Import Validity}

In our first research question we investigate to what extent \ourtool can generate tests without import errors while also avoiding distorted generations (i.e., import streaks).
The results for the two tasks and programming languages are shown in \Cref{tab:res_rq1}.
We see that the unconstrained test generation methods generate import errors accounting for 11--35\% of all instances in Java and 6--12\% in Python, with the reasoning model \gptoss yielding generally fewer errors than \llama.
In contrast, when test generation methods use \ourtool, they achieve a 0\% import error rate, across all tasks, models, and programming languages, empirically confirming that constrained decoding can provide trustworthy guarantees for import-error-free code.

The \textit{Constrained Decoding} baseline also achieves a 0\% import error rate, but generates import streaks as a by-product. 
For example, out of the 22 import errors prevented by constrained decoding (\fp generation in Java), 16 result in import streaks with the \gptoss model, while 12 out of 36 prevented errors result in import streaks with the \llama model.
In contrast, \ourtool generated one and two import streaks respectively, highlighting the value of beam constrained decoding.
This shows that \ourtool provides validity guarantees but also balances output code quality.

LIBRO's dependency resolution baseline leaves the lowest number of import errors unfixed (0--30\%) among non-constraining baselines, but still does not guarantee the 0\% rate of \ourtool.
The RAG approach leaves 31--82\% of import errors unfixed across tasks, languages, and models, suggesting that static retrieval is less effective than enforcing import validity during generation.
Finally, \textit{Agentic Repair} leaves 5--78\% of import errors unfixed, suggesting that execution feedback is not enough to guarantee the absence of import errors, and in general it is an ineffective way to impose constraints~\cite{nagy2026chopchop}.

To better understand why Agentic Repair fails, we manually analyzed the \num{58} unresolved import errors for Task 1 in Java: \num{28} with Llama and \num{30} with GPT-OSS across three runs. 
For Llama, Agentic Repair fixed the initial import error but led to a different one in \num{13} cases, repeated the same invalid import in \num{8}, fixed only a subset of the invalid imports in \num{4}, and generated no test because the accumulated context exceeded the context window \num{3}. 
For GPT-OSS, \num{10} cases involved a different invalid import, \num{1} repeated the same invalid import, \num{5} fixed only a subset, and \num{14} were context overflow cases.

We present an illustrative example using the instance \texttt{django\_12125}, where BLAST with \gptoss and no import-validity check (i.e., unconstrained) initially generated the invalid import shown in \Cref{fig:agentic_failure} (a).
In the second attempt, agentic repair avoided this invalid import statement thanks to the error message, but generated a different invalid import shown in \Cref{fig:agentic_failure} (b).
The third attempt, which included both import errors in the prompt, yielded the same invalid import as the second attempt (\Cref{fig:agentic_failure} (c)), ignoring the instructions in the prompt.
For the same instance, \ourtool generated the import statement shown in \Cref{fig:agentic_failure} (d) in the first attempt, which was valid and resulted in a correct \fp test.

\begin{tcolorbox}[boxrule=0pt,frame hidden,sharp corners,enhanced,borderline north={1pt}{0pt}{black},borderline south={1pt}{0pt}{black},boxsep=2pt,left=2pt,right=2pt,top=2.5pt,bottom=2pt]
    \textbf{Finding 1}. \ourtool eliminates all import errors across all tasks, models, and languages, guaranteeing import-error-free test generation while substantially reducing the import streaks observed with constrained decoding.
    In contrast, RAG, LIBRO, and Agentic Repair leave 45\%, 10\%,
    and 43\% of import errors unresolved overall, respectively.
\end{tcolorbox}

\begin{table*}
    \centering
    \footnotesize
    \setlength{\tabcolsep}{1.5pt}
    \renewcommand{\arraystretch}{0.8}
    \caption{Ablation study on the hyperparameters and components of \ourtool (Task: \fp test generation).}
    \begin{tabular}{clrrrrrr}
        & & \multicolumn{3}{c}{\textbf{BLAST (\llama)}} & \multicolumn{3}{c}{\textbf{BLAST (\gptoss)}} \\
        \cmidrule(lr){3-5} \cmidrule(lr){6-8}
        \textbf{Language} & \textbf{\ourtool Variation}
        & \# Import errors ($\downarrow$)
        & \# Import streaks ($\downarrow$)
        & \# Correct tests ($\uparrow$)
        & \# Import errors ($\downarrow$)
        & \# Import streaks ($\downarrow$)
        & \# Correct tests ($\uparrow$) \\
        \midrule
\multirow{5}{*}{\shortstack{Java \\ (N=103)}} & B=1 & 0 & 12 & 10 & 0 & 16 & 35 \\
 & B=2 & 0 & 5 & 10 & 0 & 2 & 36 \\
 & B=2 + Lookup (default) & 0 & 2 & 10 & 0 & 1 & 38 \\
 & B=4 & 0 & 3 & 9 & 0 & 5 & 35 \\
 & B=4 + Lookup & 0 & 1 & 9 & 0 & 2 & 37 \\
        
        \midrule
\multirow{5}{*}{\shortstack{Python \\ (N=323)}} & B=1 & 0 & 3 & 74 & 0 & 9 & 141 \\
 & B=2 & 0 & 0 & 73 & 0 & 2 & 141 \\
 & B=2 + Lookup (default) & 0 & 0 & 75 & 0 & 1 & 141 \\
 & B=4 & 0 & 1 & 73 & 0 & 2 & 140 \\
 & B=4 + Lookup & 0 & 0 & 75 & 0 & 2 & 139 \\
        \bottomrule
        \multicolumn{8}{l}{$\downarrow$ means the lower the better, $\uparrow$ means the higher the better}
    \end{tabular}
    \label{tab:res_rq3}
\end{table*}

\subsection{RQ2---Functionally Correct Tests}\label{ssec:rq2_method}
Import errors are a major failure mode of LLM-based test generation, however, import validity is a necessary but not sufficient condition for test correctness. Therefore, we investigate to what extent the import-error-free tests generated by \ourtool are also functionally correct.
A test is considered functionally correct if it is an \fp test in Task~1 and a passing test in Task~2.
We treat passing tests as successful regardless of the coverage they achieve, because \ourtool targets import errors that prevented generated tests from even executing, and it seems reasonable to believe the coverage of the correct tests is not altered compared to unconstrained generation.
Improvements in coverage depend mostly on the test bodies and assertions, which import validity does not guarantee.

\Cref{tab:res_rq2} presents the results.
Across the evaluated settings, on average 14\% of the tests that failed with an import error in unconstrained generation turned functionally correct with \ourtool, while the remaining failed with other errors analyzed below.
This led to an overall increase of 1--4 percentage points in functional correctness compared to unconstrained generation in seven of eight settings and no change in the remaining setting, with the improvement being more visible in cases where unconstrained generation leads to more import errors.
For example, when BLAST uses \gptoss and is integrated with \ourtool, it generates \num{38} \fp tests in Java vs \num{34} in unconstrained generation, an increase of four percentage points.
Furthermore, \ourtool generates more functionally correct tests than both LIBRO and RAG baselines in 7/8 and 6/8 settings respectively.

Moreover, when given the same number of repair attempts as Agentic Repair, \ourtool generates more functionally correct tests in five cases and fewer in only one case. This finding suggests that eliminating import errors in the initial attempt frees up the repair budget to address other types of errors, ultimately increasing the number of functionally correct tests. Even without any repair attempts, \ourtool outperforms Agentic Repair in three of the eight cases and matches it in two.

Differences in functional correctness between approaches are small, so confirming them with robust statistical evidence would require many repetitions. 
We executed each experiment three times, and additional repetitions would be computationally expensive: for example, the three repetitions of \gptoss on the \fp test-generation task in Java took approximately five days.
We therefore interpret functional correctness differences descriptively rather than claiming statistically significant superiority over all baselines. 
Our results provide empirical evidence of the extent to which eliminating import errors with \ourtool translates into more correct tests.

Improvements in functional correctness are naturally smaller than the improvements in import validity (RQ1), as generated tests may still fail due to other issues. 
However, \textit{eliminating import errors is beneficial even when the resulting test is not functionally correct} because the entire repair budget can then be spent on the remaining failures.
To better understand these remaining failures, we manually analyzed the 168 Python instances and 174 Java instances where \ourtool eliminated an import error but the resulting generated test was still not functionally correct across all runs. 
In Python, the top three failure modes were attribute errors (41), assertion errors (29), and type errors (17). 
In Java, type errors in the test body are the most common (97), followed by attribute errors (78), and access violation errors (21).

\begin{tcolorbox}[boxrule=0pt,frame hidden,sharp corners,enhanced,borderline north={1pt}{0pt}{black},borderline south={1pt}{0pt}{black},boxsep=2pt,left=2pt,right=2pt,top=2.5pt,bottom=2pt]
    \textbf{Finding 2}. 
    On average, \ourtool converts 14\% of import errors into functionally correct tests, increasing overall correctness by 2 percentage points (range: 1--4) compared to unconstrained generation.
    Even when eliminating import errors does not directly produce correct tests, it frees up the repair budget to address other types of errors. This leads to an average improvement in correctness over unconstrained generation of 4 percentage points (range: 2--7) in Python and 3 percentage points (range: 1--6) in Java.
\end{tcolorbox}

\subsection{RQ3---Ablation}
In RQ3 we investigate how the components and hyperparameters of \ourtool affect its performance.
To this end, we repeat the experiments of the \fp test generation task with the variations shown in \Cref{tab:res_rq3}.
The first row (\textit{B=1}) corresponds to constrained decoding~\cite{geng-etal-2023-grammar}.
We evaluate $B=2$ and $B=4$ as beam sizes for the beam-constrained decoding component, each one evaluated with and without the lookup component.
We report the number of import errors, import streaks, and functionally correct \fp tests.

\Cref{tab:res_rq3} shows that all variations lead to 0\% import error rate, indicating that \ourtool guarantees import validity regardless of its hyperparameters.
Regarding import streaks, they are drastically reduced with $B=2$, so we use this as the default value instead of the more computationally intensive $B=4$.
Finally, the lookup component either increases or leaves the number of \fp tests the same across models and tasks.
Therefore, we select the universal configuration of $B=2$ with lookup as the default.

\begin{tcolorbox}[boxrule=0pt,frame hidden,sharp corners,enhanced,borderline north={1pt}{0pt}{black},borderline south={1pt}{0pt}{black},boxsep=2pt,left=2pt,right=2pt,top=2.5pt,bottom=2pt]
    \textbf{Finding 3}. \ourtool achieves a 0\% import error rate across all evaluated configurations.
    A beam size of 2 is sufficient to mitigate import streaks, and the lookup mechanism increases or preserves the number of functionally correct tests.
\end{tcolorbox}

\begin{table*}
    \centering
    \footnotesize
    \setlength{\tabcolsep}{6.2pt}
    \renewcommand{\arraystretch}{0.8}
    \caption{Comparison between unconstrained test generation with \gptoss (row 1), \gptoss constrained with \ourtool (row 2), \gptcommercial (row 4), and a hybrid method (row 3). Bold-faced values indicate better results.}
    \begin{tabular}{llrrrrrrrr}
        & & \multicolumn{4}{c}{\textbf{\# Import Errors ($\downarrow =$ better)}} & \multicolumn{4}{c}{\textbf{\# Functionally Correct Tests ($\uparrow =$ better)}} \\
        \cmidrule(lr){3-6} \cmidrule(lr){7-10}
        & & \multicolumn{2}{c}{\textbf{Java}} & \multicolumn{2}{c}{\textbf{Python}} & \multicolumn{2}{c}{\textbf{Java}} & \multicolumn{2}{c}{\textbf{Python}} \\
        \cmidrule(lr){3-4} \cmidrule(lr){5-6} \cmidrule(lr){7-8} \cmidrule(lr){9-10}
        \textbf{Imports From} & \textbf{Body From} & \textbf{\shortstack{\fp\\(BLAST)}} & \textbf{\shortstack{Regr.\\(ChatTester)}} & \textbf{\shortstack{\fp\\(BLAST)}} & \textbf{\shortstack{Regr.\\(ChatTester)}} & \textbf{\shortstack{\fp\\(BLAST)}} & \textbf{\shortstack{Regr.\\(ChatTester)}} & \textbf{\shortstack{\fp\\(BLAST)}} & \textbf{\shortstack{Regr.\\(ChatTester)}} \\
        \midrule
GPT-OSS & GPT-OSS & 22 & 11 & 26 & 38 & 34 & 36 & 135 & 144 \\
GPT-OSS-\ourtool & GPT-OSS & \textbf{0} & \textbf{0} & \textbf{0} & \textbf{0} & 38 & 37 & 141 & 157 \\
GPT-OSS-\ourtool & \gptcommercial & 2 & 1 & \textbf{0} & \textbf{0} & \textbf{50} & \textbf{57} & \textbf{158} & \textbf{201} \\
\gptcommercial & \gptcommercial & 18 & 7 & 8 & 7 & 47 & 56 & 154 & 198 \\
        \bottomrule
        \multicolumn{10}{l}{GPT-OSS: \gptoss}
    \end{tabular}
    \label{tab:res_rq4}
\end{table*}

\subsection{RQ4---Combining with Commercial LLMs}\label{ssec:rq4_results}
RQ1--RQ3 evaluated \ourtool with local LLMs, which are often preferred due to privacy, cost, and offline use~\cite{alizadeh2026green}, but may underperform their closed-source commercial counterparts.

In this research question, we split test generation into two subtasks: import and test body generation, allowing us to evaluate hybrid approaches that combine locally constrained import generation with commercial LLM body generation.
The first column (\textit{Imports From}) of \Cref{tab:res_rq4} shows the model used to generate the imports given the task prompt.
The generated imports are then appended to the task prompt and fed to the model shown in the second column (\textit{Body From}), which generates the test body.
The rest of the columns report the number of import errors and functionally correct tests for each combination across all tasks and programming languages.

\Cref{tab:res_rq4} shows that even \gptcommercial (fourth row) produces a non-negligible number of import errors (i.e., 18/103 in Java), although fewer than \gptoss and far fewer than \llama.
\gptcommercial's import errors are proportionally more frequent on GHRB than on SWT-Bench, and one possible explanation is differing exposure to benchmark instances during training: GHRB is periodically updated with recent bugs every three months, whereas SWT-Bench is not. 
However, other factors like difference in programming language may also contribute.
Interestingly, the highest number of functionally correct tests across both tasks is achieved by combining constrained import generation with \ourtool on a local model with \gptcommercial for test body generation.
This hybrid approach results in a median of 50/103 \fp tests in Java compared to 47/103 when the whole generation is delegated to \gptcommercial.

Another finding is that in three cases GPT-5.2 deviates from our instructions to use only the valid imports provided in the prompt, rather it adds new imports leading to import errors.
This further highlights the value of constrained decoding, which enforces validity during token generation rather than relying on prompt adherence.

These results suggest that LLM-based test generation benefits from separating grounded subproblems from open-ended generation.
In particular, imports are better handled by a local model that can be constrained to provide validity guarantees, while the test body benefits from the broader reasoning capabilities of a frontier model.
More broadly, this points to a hybrid design principle for LLM-based software engineering pipelines: using guarantees where possible and reserving frontier models for subproblems that require open-ended generation.
This decoupling provides a dual benefit: higher functional correctness at a lower cost.

\begin{tcolorbox}[boxrule=0pt,frame hidden,sharp corners,enhanced,borderline north={1pt}{0pt}{black},borderline south={1pt}{0pt}{black},boxsep=2pt,left=2pt,right=2pt,top=2.5pt,bottom=2pt]
    \textbf{Finding 4}. A hybrid test generation approach where imports are generated by a small model constrained with \ourtool while test-body generation is delegated to a frontier model, generates more functionally correct tests than delegating the whole generation to the frontier model.
    The improvement is larger in the GHRB dataset where the frontier model generates more import errors initially.
\end{tcolorbox}

\begin{table}
    \centering
    \scriptsize
    \setlength{\tabcolsep}{4pt}
    \caption{Generation time ($t_{\textit{gen}}$) and end-to-end time ($t_{\textit{e2e}}$) in seconds. The mean and st.d. are reported over instances with import errors (\textit{On import errors}) and over all instances (\textit{On all instances}). 
    }
    \begin{tabular}{llrrrr}
        & & \multicolumn{2}{c}{\textbf{On import errors}} & \multicolumn{2}{c}{\textbf{On all instances}} \\
        \cmidrule(lr){3-4} \cmidrule(lr){5-6}
        \textbf{Model} & \textbf{\shortstack{Import Validity\\Approach}} & $t_{\textit{gen}}$ & $t_{\textit{e2e}}$ & $t_{\textit{gen}}$& $t_{\textit{e2e}}$ \\
        \midrule
\multirow{4}{*}{\shortstack{Llama}} & Unconstrained & 171 $\pm$ 61 & 196 $\pm$ 66 & 151 $\pm$ 79 & 169 $\pm$ 82 \\
 & \ourtool & 233 $\pm$ 218 & 256 $\pm$ 213 & 172 $\pm$ 152 & 190 $\pm$ 151 \\
        \cmidrule(lr){2-6}
 & Agentic Repair & 521 $\pm$ 153 & 595 $\pm$ 174 & 273 $\pm$ 215 & 309 $\pm$ 244 \\
 & CBDL + Repair & 537 $\pm$ 195 & 602 $\pm$ 200 & 279 $\pm$ 232 & 311 $\pm$ 254 \\
        \midrule
\multirow{4}{*}{\shortstack{GPT-OSS}} & Unconstrained & 283 $\pm$ 167 & 309 $\pm$ 169 & 205 $\pm$ 170 & 228 $\pm$ 168 \\
 & \ourtool & 350 $\pm$ 225 & 376 $\pm$ 216 & 221 $\pm$ 187 & 244 $\pm$ 185 \\
        \cmidrule(lr){2-6}
 & Agentic Repair & 929 $\pm$ 532 & 1002 $\pm$ 533 & 346 $\pm$ 369 & 381 $\pm$ 385 \\
 & CBDL + Repair & 584 $\pm$ 345 & 641 $\pm$ 351 & 275 $\pm$ 260 & 306 $\pm$ 270 \\
        \bottomrule
        \multicolumn{6}{l}{Llama: \llama, GPT-OSS: \gptoss}
    \end{tabular}
    \label{tab:overhead}
\end{table}

\subsection{RQ5---Overhead}
\ourtool introduces a performance overhead over the unconstrained generation due to (i)~the constraining algorithm (Lines~5---9 in \autoref{alg:constrained_decoding}), and (ii)~beam search.
We quantify this overhead in our fifth research question.

A test generation pipeline consists of the generation stage and the test evaluation stage that sets up the project and executing the test.
\ourtool additionally requires a one-time valid import extraction setup; since it runs only once and its runtime is negligible, we exclude it from the measurements.
Let $t_{\textit{gen}}$ denote the time for generation and $t_{\textit{eval}}$ the time for evaluation; the total time to complete the test generation task end-to-end is $t_{\textit{e2e}} = t_{\textit{gen}} + t_{\textit{eval}}$.
We measure $t_{\textit{e2e}}$ by logging the time at the beginning and end of the testing pipeline, and $t_{\textit{gen}}$ by logging the time at the beginning and end of the LLM generation.
The mean and standard deviation aggregated over all dataset instances (\textit{On all instances} column), and over all instances where \ourtool activated its constraining algorithm (\textit{On import errors}) are shown in \Cref{tab:overhead} for the \fp test generation task in Java.
For this RQ we do not reuse the reasoning trace as described in~\Cref{sssec:non_determinism} so that runtime measurements include the full reasoning and code generation cost.

For \gptoss, the mean generation time ($t_{\textit{gen}}$) over all instances is \num{16}s higher for \ourtool compared to unconstrained generation (\num{221}s vs \num{205}s) while for \llama \ourtool is \num{21}s slower on average (\num{172}s vs \num{151}s).
However, we see that Agentic Repair is slower than a matched-budget version of \ourtool in terms of maximum number of LLM queries, for example taking \num{929}s vs \num{584}s to repair import errors with \gptoss.
We find that this is partly because Agentic Repair uses the second repair iteration more frequently than \ourtool, whose initial attempt already eliminates import errors: on average \num{11.7} instances require a second repair iteration in \ourtool and \num{14} in Agentic Repair.
Additionally, each individual repair attempt takes longer on average with Agentic Repair (\num{347}s) than \ourtool equipped with repair on top (\num{224}s), possibly because more errors are simultaneously present in the repair prompts.
For \llama, Agentic Repair achieves slightly lower generation time despite requiring more repair attempts in two instances (\num{537}s vs \num{521}s).

\begin{tcolorbox}[boxrule=0pt,frame hidden,sharp corners,enhanced,borderline north={1pt}{0pt}{black},borderline south={1pt}{0pt}{black},boxsep=2pt,left=2pt,right=2pt,top=2.5pt,bottom=2pt]
    \textbf{Finding 5}. \ourtool introduces an average overhead of 16--21 seconds per test over unconstrained generation. However, Agentic Repair is slower than repairing attempts initiated with \ourtool given equal number of maximum repair attempts with \gptoss (\num{929}s vs \num{584}s) and slightly faster with \llama (\num{521}s vs \num{537}s) in generation time.
\end{tcolorbox}

\section{Discussion}

\subsection{Implications for Practitioners}
\ourtool eliminates a major failure cause in automated test generation, enabling more cost-effective pipelines that focus their budget on remaining failures. 
Furthermore, the hybrid approach combining locally constrained import generation with remote test body generation by a frontier model yields more functionally correct tests than passing the whole generation task to a frontier LLM.
Organizations using automated test generation pipelines, such as Mozilla~\cite{blast} or Meta~\cite{meta}, can generate more functionally correct tests at a lower cost.

\subsection{Implications for Researchers}
Despite LLMs' success in software engineering (SE), their unpredictability and lack of guarantees remain open challenges.
While most techniques rely on prompt engineering or agentic feedback loops, our results show that constrained decoding can provide validity guarantees while maintaining generation quality.
We view this work as a step toward guarantee-driven LLMs for SE and envision future applications of constrained decoding to enforce other properties (e.g., correct API usage or project-specific rules). 
Moreover, since prior work identifies import errors as a major failure mode, eliminating them opens the door for future research on deeper improvements in test quality and semantic correctness.

\subsection{Threats to Validity}
To mitigate the effect of LLMs' stochastic nature on the validity of our results, we repeat each generation three times~\cite{ouyang2025empiricalnondeterminism}.
We also conduct experiments on two test generation tasks, two programming languages, two underlying test generation methods, and three LLMs, to mitigate generalizability threats.

Another threat is memorization: since the LLMs' training data is not public, some benchmark instances may overlap with it.
We mitigate this by using GHRB, which is updated every three months with new bugs.
Moreover, substantial memorization would result in fewer import errors, as models would simply reproduce memorized imports, so \ourtool's usefulness would likely be even greater in a memorization-free setting.

\section{Related Work}\label{sec:related_work}

\subsection{Constrained Decoding}\label{ssec:rel_work_constrained}
Prior work has explored constrained decoding for structured generation. GCD~\cite{geng-etal-2023-grammar}, GAD~\cite{grammar_alinged_decoding}, and XGrammar~\cite{dong2024xgrammar} enforce grammar constraints during non-code generation. 
For code generation, SynCode~\cite{ugare2025syncode} proposes syntax-constrained generation, while type-constrained generation has been studied through monitor-guided decoding~\cite{NEURIPS2023_662b1774} and automata-based constraints~\cite{muendler2025typeconstrained}. 
Prior work has shown that constrained decoding can enforce grammar constraints, however, enforcing import validity does not necessarily lead to useful test generations due to import streaks, which \ourtool addresses through beam search and lookup.
To the best of our knowledge, we are also the first to show that a constrained decoding approach for import validity outperforms strong baselines (including agentic repair) and that delegating import generation to a locally constrained model improves the correctness of tests generated by a frontier model.

\subsection{LLM-Based Test Generation}\label{ssec:rel_work_testgen}
Several LLM-based approaches have been proposed for \fp test generation, including LIBRO~\cite{libro}, AutoTDD~\cite{tddbench}, SWT-Agent~\cite{swtbench}, Issue2Test~\cite{issue2test}, e-Otter++~\cite{otter}, and BLAST~\cite{blast}. 
Similarly, regression test generation methods include ChatUniTest~\cite{chatunittest}, ChatTester~\cite{chattester}, TestPilot~\cite{testpilot}, and CoverUp~\cite{coverup}. 
These approaches primarily rely on prompt engineering, iterative feedback, or agentic repair loops.
We show that \ourtool can be integrated with existing test generation approaches, and, by guaranteeing import validity, it improves the functional correctness of the generated tests.

Prior work has also addressed missing imports in incomplete code, e.g., from Stack Overflow. 
SnR~\cite{snr} infers dependencies from a knowledge base while ZS4C~\cite{zs4c} uses a repair loop to synthesize compilable code from incomplete snippets.
Instead, \ourtool enforces validity before repair becomes necessary.

\section{Conclusion}
We introduce \ourtool, a technique that enforces valid imports in LLM-based test generation, while mitigating low-quality outputs resulting from constrained decoding.
\ourtool thus eliminates import errors, one of the most common errors in LLM-based test generation and in doing so it also leads to more correct tests.
Furthermore, generating imports with a local \ourtool-equipped LLM and the test body with a frontier model improves functional correctness over the frontier model alone. 
These findings demonstrate the value of constraint-based techniques within LLM-based software engineering pipelines.

\bibliographystyle{IEEEtran}
\bibliography{references}

\end{document}